\documentclass[12pt,preprint]{aastex}
\usepackage{natbib}
\begin{document}

\title{In-spiraling Clumps in Blue Compact Dwarf Galaxies}

\author{Bruce G. Elmegreen}
\affil{IBM Research Division, T.J. Watson Research Center, 1101
Kitchawan Road, Yorktown Heights, NY 10598} \email{bge@watson.ibm.com}

\author{Hong-Xin Zhang}

\affil{Lowell Observatory, 1400 West Mars Hill Road, Flagstaff, Arizona
86001 USA; Purple Mountain Observatory, Chinese Academy of Sciences, 2
West Beijing Road, Nanjing 210008 China; Graduate School of the Chinese
Academy of Sciences, Beijing 100080, China}

\author{Deidre A. Hunter}

\affil{Lowell Observatory, 1400 West Mars Hill Road, Flagstaff, Arizona
86001 USA}

\begin{abstract}
Giant star-formation clumps in dwarf irregular galaxies can have masses
exceeding a few percent of the galaxy mass enclosed inside their
orbital radii. They can produce sufficient torques on dark matter halo
particles, halo stars, and the surrounding disk to lose their angular
momentum and spiral into the central region in 1 Gyr. Pairs of giant
clumps with similarly large relative masses can interact and exchange
angular momentum to the same degree. The result of this angular
momentum loss is a growing central concentration of old stars, gas, and
star formation that can produce a long-lived starburst in the inner
region, identified with the BCD phase. This central concentration is
proposed to be analogous to the bulge in a young spiral galaxy.
Observations of star complexes in five local BCDs confirm
the relatively large clump masses that are expected for this process.
The observed clumps also seem to contain old field stars, even after
background light subtraction, in which case the clumps may be
long-lived. The two examples with clumps closest to the center have the
largest relative clump masses and the greatest contributions from old
stars. An additional indication that the dense central regions of BCDs
are like bulges is the high ratio of the inner disk scale height to the
scale length, which is comparable to 1 for four of the galaxies.
\end{abstract}


\section{Introduction}

Blue Compact Dwarfs (BCDs) are small galaxies with intense
emission-lines from starburst HII regions in their central regions
\citep{sargent70}.  They are gas-rich like other dwarf irregulars
\citep{cham77,gordon81,thuan81}, but much more centrally concentrated
in stars \citep{noeske03,hunter06}, star formation
\citep{heller00,hunter04}, and gas \citep{taylor94,vanzee98b}. This
concentration suggests that gas inflow following angular momentum loss
led to enhanced star formation in a dense and gravitationally unstable
central disk \citep{taylor94,papaderos96,vanzee01}.  Gas loss from the
outer parts also produces a shrinking radius for star formation, as
observed in dwarfs \citep{zhang11,koleva11}.

Individual star-forming regions in BCDs are relatively large, giving
the galaxies a clumpy, irregular appearance in H$\alpha$
\citep[e.g.][]{kunth88,cairos09b} and FUV \citep{thuan97}. Clump
emission lines are supersonic and apparently virialized
\citep{terlevich81,firpo11}, which implies the clumps could last for
several internal crossing times. Low velocity shear
\citep{thuan99,vanzee01,ramya11} and resolved stellar population
studies \citep{dohm98} also suggest the clumps could be long-lived, 100
Myr or more.  \cite{mcquinn10} and \cite{zhang11} suggest the starburst
itself can last for $\sim1$ Gyr.

Surrounding many BCDs are pools of HI, sometimes as large as 4 or more
optical radii \citep{brinks88,taylor96,vanzee98b,putman98,pustilnik01,
hoffman03}. The peripheral HI is often not simply rotating
\citep{vanzee98b} or even in a disk \citep{pustilnik97,pustilnik01}.
This peripheral gas led to suggestions about cloud impacts
\citep{gordon81}, weak interactions
\citep{brinks88,pustilnik02,bravo04}, merging \citep{bekki08}, and
tidal effects \citep{vanzee98b,pustilnik01} in efforts to explain the
high star formation rates. Alternatively, the HI could be vestigial
streams or pools of cosmological accretion
\citep{taylor93,thuan97b,vanzee98a,wilcots98}, like the streams modeled
for higher-mass galaxies \citep[e.g.,][]{ceverino10}.

BCDs resemble young galaxies in many respects
\citep{e09b,izotov11,griffith11}. They are gas-rich, low-metallicity
\citep{izotov01}, relatively turbulent
\citep{vanzee98b,vanzee01,silich02,garcia08}, and highly clumped with
locally intense star formation. They tend to lie at the edges of
galactic clusters or in voids with only low-mass neighbors, and where
harsh environmental effects like ram pressure stripping are minimal
\citep{grogin00,drinkwater01,pustilnik02}. Those with the lowest
metallicity tend to be relatively young, having formed most of their
stars within the last few Gyrs
\citep{searle72,noeske00,johnson00,fricke01,papaderos08}. BCDs with
less extreme metallicities tend to have relatively more old stars
\citep[e.g.,][]{thuan83,loose86b,crone02,caon05,cairos07,
cairos09a,cairos09b,zhao11,zhang11}. In the most low-metal BCDs, there
is little evidence for stars older than $\sim200-500$ Myr; these
include SBS 1415+437 \citep{thuan99,guseva03a}, SBS 1129+576
\citep{guseva03b}, and I Zw 18 \citep{papaderos02}.

Specific examples of BCDs illustrate these points. I Zw 18 is in many
respects a morphologically young system. It has two giant star-forming
regions inside a kpc-scale blue continuum of stars, ionized emission
with an overall exponential profile \citep{papaderos02}, and an
extensive HI envelope \citep{vanzee98a}. Inside each region the star
formation is widely distributed \citep{hunt05}. The rotation curve is
flat but steeply rising in the inner part, where baryons may dominate
dark matter \citep{lelli11}. Radial motions in the disk of $\sim15$ km
s$^{-1}$ suggest a major disturbance, which \cite{lelli11} suggest is
tidal because there is a dwarf companion galaxy and elongated
peripheral HI gas. The rotation curve gives a mass of $10^8\;M_\odot$
in which $\sim70\%$ is neutral gas \citep{ramos11}. It has very low
metallicity \citep[2\%-3\% solar,][]{izotov01} even though old stars
are present. \cite{recchi04} and others have considered the selective
removal of metals by winds.

Another well-studied example is VII Zw 403, which is among the nearest
BCDs to the sun, having a distance of 4.5 Mpc \citep{lynds98}. VII Zw
403 has a half-dozen big clumps and many H$\alpha$ filaments from
superbubbles \citep{lozinskaya06} in the midst of a smooth elliptical
background of old red giant stars \citep{schulte98}. The dynamical mass
is $2\times10^8\;M_\odot$, with approximately 20\% in HI
\citep{thuan81}. The rotation speed is only $\sim15$ km s$^{-1}$
\citep{simpson11} and the metallicity is 5\% solar
\citep{martin97,izotov97}. \cite{lynds98} used resolved stellar
populations to date a major star burst to 600 Myr ago, when the star
formation rate was $\sim30$ times higher than it is today. The current
burst produced $\sim10^6\:M_\odot$ within the last 10 Myr
\citep{silich02}.

The most intense star formation in BCDs can occupy very compact regions
with extreme densities and local formation rates. SBS 0335-052 is a
pair of extremely young interacting dwarf galaxies
\citep{pustilnik01,ekta09} without much of an underlying old population
\citep{papaderos98}, and with a metallicity of 2.5\%--4\% solar
\citep{izotov09a,peimbert10}. There are 6 super star clusters
\citep{thuan97}, of which two, within $\sim200$ pc of each other, have
extremely intense star formation. \cite{hunt05} and \cite{johnson09}
found radio free-free absorption and a very high emission measure where
the electron density is $\sim10^3$--$10^4$ cm$^{-2}$, the star
formation rate is $\sim1\;M_\odot$ yr$^{-1}$, and the excitation comes
from the equivalent of $\sim10^4$ O7 stars.

Here we propose that central accretion and long-lived starbursts in
some BCDs arise from gravity-driven motions and torques produced by
clump formation, clump dynamical friction, and clump interactions --
the same processes that could make bulges in larger galaxies
\citep[e.g.][]{noguchi99,immeli04,bournaud07}. BCDs have steep stellar
profiles in the inner 500 pc that are exponential \citep{hunter06} or
deVaucouleur's \citep{doublier99}, as in the bulges of earlier Hubble
types. Such high central concentrations require baryonic mass inflow
and significant angular momentum redistribution in the disk.  Much of
this inflow could have occurred when BCDs were young, but some of
today's BCDs still look dynamically young even if there are old stars,
and significant inflow could be occurring now.

Our emphasis differs from that in \cite{governato10}, where simulations
of dwarf galaxies highlight the removal of gas in order to avoid
central concentrations. In these simulations, diffusion, torques, and
pressure-driven inflows return some of this gas to the center
\citep[e.g.,][]{recchi06,dal10}, only to have it removed again by the
next starburst, cycling in and out many times
\citep[e.g.][]{stinson07,revaz09}. The result is a bulge-free late-type
galaxy and a time-changing central potential that converts a primordial
dark matter cusp into a more uniform dark matter core \citep{read05}.
The degree of this conversion varies for different simulations
\citep{ogiya11,oh11}. \cite{alard11} note that the least evolved
galaxies, having the highest relative gas abundances, tend to have the
steepest inner density profiles, supporting the idea that gas recycling
and star formation make the inner profiles shallow over time.  These
observations could imply that some BCDs still have steep central dark
matter profiles, if these galaxies are relatively young.
\cite{delpopolo11} also model low-mass galaxies and suggest that tidal
torques and the baryon fraction before collapse influence the central
density profile of dark matter.  Steeper central profiles are predicted
to occur in more remote galaxies and in those with higher dark matter
fractions; BCDs could be in this category too \citep{grogin00}.

BCDs are unusual in having both a central concentration and a high gas
abundance. This combination also appeared in massive galaxies at
redshift $z\sim2$ \citep{e09a}. Observations at intermediate-to-high
redshift indicate that Hubble types arise mostly since $z\sim1$
\citep{papovich05,bundy06}. The morphology seems to depend on dynamical
maturity. High-mass disks like Hubble type Sa tend to be higher density
than low-mass disks like Hubble type Sd \citep{roberts94}. Thus, Sa's
evolve more quickly to a centrally concentrated state with a low gas
fraction and a low specific star formation rate
\citep{sandage86,zhang07}.  If this trend of increasing central
concentration and decreasing specific star formation rate continues
into the future for low-mass galaxies, then some dwarf Irregulars might
also evolve to a centrally concentrated state with little remaining
gas. Mass loss from winds and supernovae in the low potential well of
the dwarf could prevent such a central concentration however, depending
on the relative rates of inflow from torques and outflow from winds.
Those with dominant inflows could go through the BCD phase, as
discussed here. Strong inflow depends on the presence of relatively
massive clumps or tidal arms.

In what follows, we estimate the accretion time of a clumpy disk from
dynamical friction (Sect. 2), and then consider whether observed clumps
and other irregularities in BCDs are massive enough to drive
significant disk evolution on a Gyr time scale (Sect. 3). A summary is
in Section 4.

\section{Clump Accretion in BCDs}
\subsection{The Case with a few Giant Clumps}

The process of clump drag and interaction leading to coalescence in the
center of a galaxy has been illustrated with detailed simulations of
high redshift galaxies \citep{elmegreen08,ceverino10}. The clumps in
these simulations formed spontaneously in a turbulent disk and had
masses of about 5\% of the total galaxy mass. Migration to the center
took only a few orbit times ($\sim0.5$ Gyr). Clump destruction by star
formation feedback \citep{genel10} does not stop the torques and
accretion if each destroyed clump is replaced by a new one. This
replacement is likely as long as the conditions for forming the first
clumps, such as high gas fractions and turbulent speeds, are still
present. A low ratio of turbulent speed to orbit speed would stop this
process, because then the clumps that form by gravitational
instabilities are relatively low-mass and produce proportionally weaker
torques.  The biggest star-forming regions in the Milky Way are only
$\sim10^{-3}$ times the disk mass and should have little tendency to
move to the center.

Dynamical friction and clump torques are important if the ratio of the
disk Jeans mass to the galaxy mass is more than a few percent. This
ratio scales with the square of the ratio of the gas velocity
dispersion to the rotation speed. In high redshift galaxies, the
rotation speed is normal for a massive disk but the dispersion is
abnormally high \citep[][and references therein]{erb06,forster11}
making the ratio high. In local dwarf Irregulars, the gas dispersion is
normal for local galaxies, $\sim10$ km s$^{-1}$, but the rotation speed
is low, $\sim50$ km s$^{-1}$ or less. In both cases the ratio of speeds
is high and the clumps that form by gravitational instabilities are
massive compared to the disk. The same processes of massive clump
formation and angular momentum exchange should happen in high redshift
galaxies and local gassy dwarfs because both have relatively large
velocity dispersions compared to rotation speeds.

The timescale for dynamical friction between an orbiting clump of mass
$M_{\rm c}$ in the disk and non-rotating dark matter particles or stars
in a halo is $v\left(dv/dt\right)^{-1}$ where
\begin{equation} {{dv}\over{dt}}= {{ 4 \pi \ln\Lambda G^2M_{\rm c}
\rho}\over v^2} \left( {\rm erf}\left[X\right] - {{ 2
X}\over{\pi^{1/2}}}e^{-X^2} \right)={{ 4 \pi \ln\Lambda G^2M_{\rm
c}\rho\xi}\over v^2} \label{dvdt}\end{equation}\citep{binney08}. Here,
$v$ is the clump orbital speed, $X=v/(2^{1/2}\sigma)$ for halo 3D
velocity dispersion $\sigma$, $\rho$ is the halo density, $\ln\Lambda$
is the coulomb factor, and $\xi$ is the quantity in parentheses. This
formula assumes that the clump is a self-gravitating object surrounded
by a uniform density of low-mass field stars or dark matter particles
that have a Maxwellian velocity distribution function.

It is convenient from an observational point of view to write the local
rotation speed as a power of the local radius, $v(r)\propto r^\beta$,
since $\beta$ comes from the rotation curve.  Starting with
$\rho(r)=\rho_0r^{-\alpha}$ and $v(r)^2=GM_{\rm dyn}(r)/r$, we get
$v(r)^2/(4\pi\rho[r] Gr^2)=1/(3-\alpha)=1/(1+2\beta)$. Then the
dynamical friction time, $v/(dv/dt)$, in units of the dynamical orbit
time, $r/v$, is
\begin{equation}
T(r) \equiv{{v^2}\over{r(dv/dt)}}={{1}\over{\ln\Lambda
\xi(1+2\beta)}}\times {{M_{\rm dyn}(r)}\over{M_{\rm c}}}=T_0(r){{M_{\rm
dyn}(r)}\over{M_{\rm c}}}\label{time}\end{equation} where $M_{\rm dyn}$
is the galaxy dynamical mass enclosed within the orbital radius of the
clump.

Dwarf galaxies have nearly solid body rotation in the inner parts
\citep{swaters02}. BCDs can have steeply rising rotation curves in the dense
inner regions, and flatter rotation curves beyond that
\cite[e.g.,][]{vanzee98b,lelli11}. The BCDs we consider in Section 3 have
approximately-linear rising rotation curves in the vicinity of the giant
clumps, and some have flat rotation curves beyond that \citep[e.g., NGC
2366;][]{thuan04}. Thus for the main starburst regions we can take
$\beta\sim1$ or slightly less. If $v\sim\sigma$, then $\xi\sim0.20$. In that
case, the dimensionless time coefficient in equation (\ref{time}) is
$T_0=0.56$ for typical $\ln\Lambda=3$ (see below). For a clump at $r=0.5$ kpc
orbiting with $v=10$ km s$^{-1}$, $r/v=49$ Myr and the dynamical friction
time in physical units is $27M_{\rm dyn}/M_{\rm c}$ Myr. This means that a
clump with a mass greater than 2.7 percent of the enclosed galaxy mass has
$T<1$ Gyr.

Note that $M_{\rm dyn}$ decreases with radius, while the normalization
quantity, $r/v$, is constant for $\beta\sim1$. Thus the timescale gets
smaller as the clump moves in.  Writing the rate of change of clump angular
momentum as $dL/dt=M_{\rm c}r(dv/dt)$ for frictional deceleration in the
azimuthal direction $dv/dt$, and setting this equal to $M_{\rm c}v(dr/dt)$
for circular speed $v$ and radial drift speed $dr/dt<<v$, we get
$dr/dt=v/T(r)$. For $\beta=1$, $v\propto r$ and $M_{\rm dyn}\propto r^3$ so
$T(r)\propto r^3$ if $\Lambda$ and $\xi$ are constant. Then, the time to
reach the center is 1/3 the instantaneous $T$ in equation \ref{time}.  If
$\beta=1/2$, then $v\propto r^{1/2}$ and $M_{\rm dyn}\propto r^2$ so
$dr/dt\propto r^{-3/2}$ and it takes $0.4T$ to reach the center.

\begin{figure}
\centering
\includegraphics[width=6.5in]{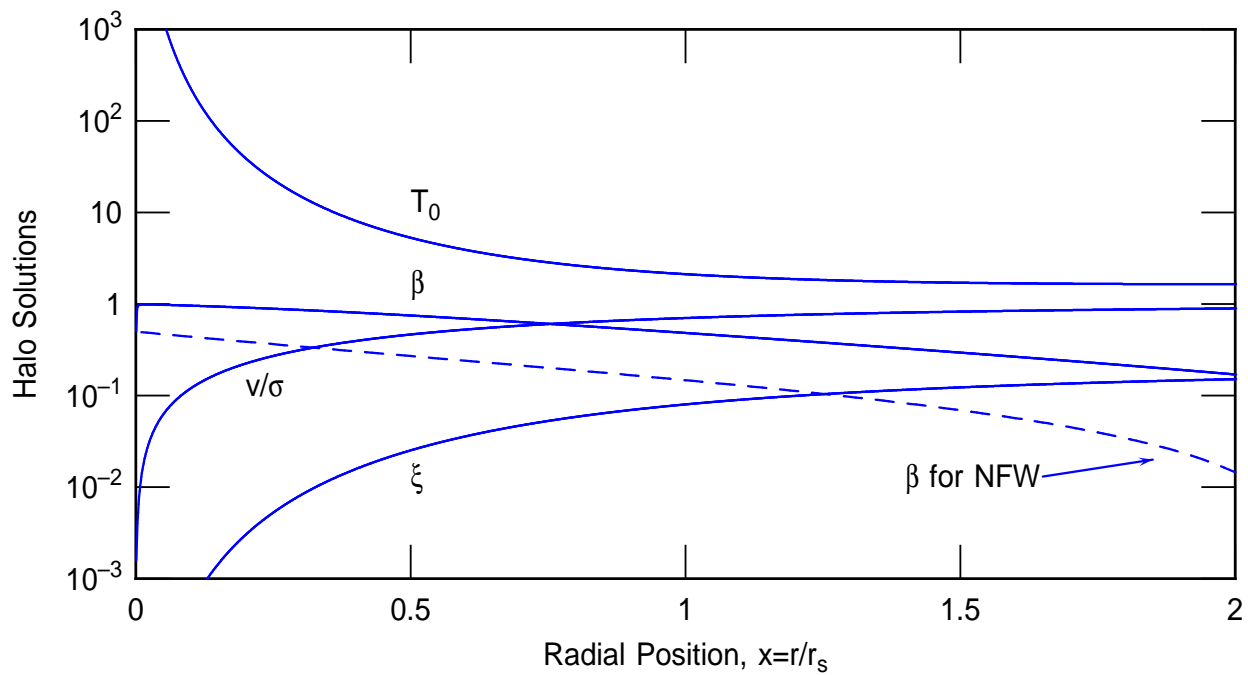}
\caption{Solutions to various parameters connected with the Burkert
(1995) dark matter density profile, which has a constant-density core;
$\xi$ is the dynamical friction parameter in the parentheses of
equation \ref{dvdt}, $v$ is the rotation speed, $\sigma$ is the 3D
velocity dispersion, $\beta$ is the slope of the rotation curve, and
$T_0$ is the prefactor in equation \ref{time}. The dashed line shows
$\beta$ for a Navarro et al. (1996) profile. }\label{hunter_bcd}\end{figure}

One uncertainty in this result is the ratio of the disk orbit speed to
the halo velocity dispersion, which enters into $\xi$.  This ratio
depends on whether the halo has a core or a cusp, and on the nature of
the core. We consider two extreme cases: a \cite{burkert95} halo
density profile in the case of a core, and an NFW \citep{navarro96}
profile for a cusp. For the Burkert profile, $\rho(x)=\rho_{\rm
s}\left([1+x][1+x^2]\right)^{-1}$ with $x=r/r_{\rm s}$ and scale
factors $\rho_{\rm s}$ and $r_{\rm s}$. We use this with the equation
of hydrostatic equilibrium in the radial direction to determine $\beta$
and $v/\sigma$ as functions of $x$. Hydrostatic equilibrium implies
$dP/dr=-GM(r)\rho(r)/r^2$ where $P=\rho\sigma_{1D}^2$ for 1D dispersion
$\sigma_{1D}=\sigma/3^{1/2}$ and $M(r)=\int_0^r 4\pi r^2\rho(r)dr$. We
assume the boundary condition $P\sim0$ and $\sigma\sim$constant at the
edge of the halo, which is taken to be where $\rho=10^{-4}\rho_0$.
Figure \ref{hunter_bcd} shows $v/\sigma$, $\xi$, $\beta$, and $T_0$ as
functions of position $x$ (determined by numerical integration).
$\beta\sim1$ for a solid body rotation, so $x$ must be small in the
visible part of the disk. For example, $x=0.54$ at the half-density
point, where $\rho=0.5\rho_0$, and there $v/\sigma=0.49$,
$\xi\sim0.03$, $\beta=0.72$, and $T_0\sim4.6$. For $r=0.5$ kpc and
$v\sim10$ km s$^{-1}$, the friction time is then $rT_0/v=220M_{\rm
dyn}/M_{\rm c}$ Myr. This implies that a clump with a relative mass of
$M_{\rm c}/M_{\rm dyn}=5$\% takes $\sim1.5$ Gyr to spiral in,
considering the factor of $1/3$ that accounts for a decreasing $M_{\rm
dyn}$ with radius, as discussed above.

The NFW dark matter profile is a little faster. For this,
$\rho=\rho_{\rm s}\left(x[1+x]^2\right)^{-1}$ with $x=r/r_{\rm s}$
again. There is a density singularity at the center that produces a
logarithmic divergence of a quantity like pressure if the equation of
hydrostatic equilibrium is considered (because $M(r)\rho/r^2\sim1/r$
near the center). We consider instead that the halo 3D velocity
dispersion is comparable to the rotation speed \citep{navarro96},
$v\sim\sigma$, which gives $\xi=0.20$.  Writing for the galaxy mass
$M_{\rm dyn}(x)=4\pi r_{\rm s}^3\rho_{\rm s}{\cal M}(x)$, where ${\cal
M}(x)=\ln(1+x)-x/(1+x)$, the slope of the rotation curve is now given
by $1+2\beta=x^2/({\cal M}(x)(1+x)^2)$. Figure \ref{hunter_bcd} shows
$\beta$ as a dashed line. We see $\beta\sim0.5$ for small $x$ and then
$T_0\sim0.83$ with $\xi=0.2$. This is a smaller prefactor than for a
cored halo because of the higher $v/\sigma$. The dynamical friction
time for an NFW halo is $\sim 41M_{\rm dyn}/M_{\rm c}$ Myr with $r=0.5$
kpc and $v\sim10$ km s$^{-1}$. Clumps with $M_{\rm c}/M_{\rm
dyn}>1.6$\% take $<1$ Gyr to spiral in, considering the factor of $2/5$
that accounts for a decreasing $M_{\rm dyn}$ with radius when $\beta=0.5$.

Dwarf galaxies do not appear to have NFW halos at the present time
\citep[e.g.,][]{zackrisson06,eymeren09,kuzio09,amor12}, although it is
difficult to be certain \citep{spekkens05,valenzuela07} and it is not yet
known whether BCDs differ from other dwarfs in this regard. An important
point for dynamical friction is the relative velocity dispersion of the field
particles, $\sigma/V$, which enters into $\xi$ as shown above. The validity
of the Chandresekhar formula is also a question. \cite{goerdt10} modeled
sinking massive objects like what we consider here and showed that a central
cusp turns into a core inside the radius where the sinking mass equals the
enclosed dark halo mass. Further sinking in their model stalled at this
radius because of a decrease in dynamical friction in the core. This result
was also found by \cite{read06} and others. In an detailed study of dynamical
friction in cored galaxies, \cite{inoue11} explained the loss of frictional
forces as a result of orbit resonances that appear when the orbit time is
independent of radius, as is the case for a constant central density. Another
limitation is that after a bulge forms, tidal forces from the bulge can rip
apart remaining clumps and prevent them from reaching the center
\citep{elmegreen08}. These considerations make it plausible that in some BCDs
today, giant clumps come in from larger radii and then stall at the edge of a
dark matter core, building up the central region from inside out. Accretion
to the center could have occurred more readily when the galaxies were young
and the dark matter profiles were more cuspy.

\cite{sanchez06} considered a similar situation with in-spiraling
globular clusters in dwarf galaxies. They concluded that dwarfs should
have dark matter cores rather than cusps so that the globular clusters
stall midway in the disk where they can still be seen.  Galaxies with
giant nuclear star clusters, however, may have allowed their disk
clusters to reach the center \citep{boker10,antonini11,hartmann11}. The
resemblance of these galaxies to BCDs is compelling if we allow for a
difference in the mass and size of the disk clumps that form by star
formation: BCDs have relatively massive and large disk clumps that
could spiral in to form massive and large central cores, while normal
dwarfs and galaxies have relatively small star clusters, which could
spiral in to form small nuclear clusters.

\subsection{Clumps with a Power-Law Mass Function}

As mentioned above, individual clumps need not survive the full trip to
the center to drive accretion if new clumps replace dispersed clumps in
a steady state. Similarly, there need not even be a single giant clump.
Any irregularities moving through a slower rotating halo will have
dynamical friction drag, and the total torque on the medium will depend
on the mean squared mass of those irregularities. Consider the
equations of disk accretion starting with the continuity equation in
two dimensions \citep{pringle81}:
\begin{equation}
{{d}\over{dt}}\Sigma+r^{-1}{{d}\over{dr}}r\Sigma v_{\rm
r}=0;\end{equation} the surface density is $\Sigma$ and the radial
drift velocity is $v_{\rm r}$.  The torque equation is
\begin{equation}
{{d}\over{dt}}r\Sigma  v +r^{-1}{{d}\over{dr}}r\Sigma v_{\rm r}r v =
{\rm torque/area}\end{equation} where $v$ is again the azimuthal speed.
For a viscous disk, the torque per unit area is $R^{-1}dG/dR$ where
$G=\nu\Sigma Ar^2$ for viscous coefficient $\nu$ (comparable to the
product of the clump mean free path and the rms speed), and Oort
rotation constant $A$, which is the rate of shear. We are not concerned
with viscosity in this paper because $A$ is small for dwarf galaxies
with little shear; $\nu$ is also usually small compared to dynamical
torques in galaxy disks \citep[however, see][]{wang09}. Here we
consider dynamical friction between orbiting clumps in the disk and the
halo and other parts of the disk. If an annular area has $N$ clumps of
mass $M_{\rm c}$, then the total torque exerted on these clumps is
$NM_{\rm c}rdv/dt$ for deceleration in the azimuthal direction $dv/dt$
from equation (\ref{dvdt}).  This quantity depends on the mean squared
clump mass because $dv/dt$ depends on clump mass. For clump mass
function $dn(M_{\rm c})/dM_{\rm c}\propto M_{\rm c}^{-\delta}$,
\begin{equation}
<NM_{\rm c}^2>={{2-\delta}\over{3-\delta}}M_{\rm c,max}<NM_{\rm
c}>=fM_{\rm c,max}<NM_{\rm c}>.\end{equation} The prefactor $f$ ranges
between $f=0.33$ at $\delta=1.5$ and $f=0.06$ at $\delta=2$ \citep[for
$\delta$, see e.g.][]{heithausen98}. In the latter case, the integral
over $M^2n(M)$ gives $\ln (M_{\rm c,max}/M_{\rm c,min})$, whose value
is $\sim16$ for typical $M_{\rm c,max}\sim10^7\;M_\odot$ and $M_{\rm
c,min}\sim1\;M_\odot$. Because $<NM_{\rm c}>/{\rm Area}=\Sigma$, we
have $<NM_{\rm c}^2>/{\rm Area}=f\Sigma M_{\rm c,max}$. Then the
torque/area for the above halo model becomes $f\Sigma
rdv/dt=\Sigma\gamma/r$ where $\gamma=(1+2\beta)f\ln\Lambda GM_{\rm
c,max}\xi$. For a fixed galactic potential ($v$ independent of time),
the torque equation is now
\begin{equation}
r^2v{{d\Sigma}\over{dt}}+ {{d}\over{dr}} r\Sigma v_{\rm r} r v =
-\Sigma \gamma.\end{equation} We can simplify this by writing
$\mu=r\Sigma v_{\rm r}$ and noting that $(d/dr)\mu r v= \mu(d/dr)rv+ rv
d\mu/dr = \mu(d/dr)rv -r^2v (d\Sigma/dt)$ using the continuity
equation. Then $d\Sigma/dt$ cancels in the torque equation and we get
\begin{equation}
r\Sigma v_{\rm r}{{drv}\over{dr}}=-\Sigma\gamma . \end{equation} This may be
solved for $v_{\rm r}$ since everything else is a known function of $r$, and
then the result can be put into the continuity equation to get $d\Sigma/dt$.
Note that $drv/dr=(1+\beta)rv$; for the other radial derivative, we set
$d/dr\sim1/r$. The resultant normalized accretion timescale is
\begin{equation}
T={{v\Sigma}\over{rd\Sigma/dt}}\sim{{(1+\beta)}\over{f\ln\Lambda
\xi(1+2\beta)}}\times {{M_{\rm dyn}(x)}\over{M_{\rm
c,max}}}=T_1(x){{M_{\rm dyn}(x)}\over{M_{\rm c,max}}}.
\end{equation}
This time is larger than before by the ratio $(1+\beta)/f$. Setting
$\beta=1$, $\ln\Lambda=3$ and $\xi\sim0.03$ for a Burkert core, and taking
$f\sim0.1$, we get $T_1=74$ and a physical accretion time of $1.2M_{\rm
dyn}/M_{\rm c,max}$ Gyr for $r=0.5$ kpc, $v=10$ km s$^{-1}$ with the factor
$1/3$ to account for a decreasing $M_{\rm dyn}$ with radius. For a NFW core
with $\beta=0.5$ and $\xi=0.2$, $T_1=12.5$ and the accretion time is
$240M_{\rm dyn}/M_{\rm c,max}$ Myr for $r=0.5$ kpc, $v=10$ km s$^{-1}$ with
the factor $2/5$. Now we see that it takes about 1 Gyr for 10\% of the ISM to
accrete to the center from the inner half-kpc if the largest cloud in a power
law distribution of cloud masses is 12\% and 2.4\% of the enclosed galaxy
mass for the Burkert and NFW profiles, respectively. If the largest cloud has
a mass much larger than the extrapolation of a power law distribution from
the other clouds, then the previous analysis for a single cloud applies.

\subsection{Clump-Clump Interactions}

Clump-clump interactions can also drive accretion by direct gravitational
forces. The acceleration on one clump by another clump is $GM_{\rm c}/\Delta
r^2$ for separation $\Delta r$.  This acceleration cumulatively distorts the
clump's motion until its velocity has changed significantly.  The timescale
for this change is $v$ divided by the acceleration, and in units of the orbit
time, it is $v^2/r$ divided by the acceleration. Writing $v^2/r=GM_{\rm
dyn}/r^2$, the normalized interaction time becomes $(\Delta r/r)^2(M_{\rm
dyn}/M_{\rm c})$. For big clumps, $\Delta r\sim r$, and the normalized
interaction time is just the ratio of masses.  That means $T_0\sim1$ in an
equation like (\ref{time}), and the accretion time is $\sim 49M_{\rm
dyn}/M_{\rm c}$ Myr for $r=0.5$ kpc and $v\sim10$ km s$^{-1}$.

Other torques will contribute to the inflow of gas, so $T$ is an upper
limit based on dynamical friction with the halo. The disk also will
produce a torque if there is shear, because then the clump will drive a
spiral wake \citep{julian66} and the wake will have its own torque that
drives mass inward \citep{lyndenbell72}.  Small galaxies tend to have
little shear, however.

\section{Observations of Clump Properties in Dwarf Irregulars and BCD Galaxies}

\subsection{Clump Mass Fractions and Accretion Times}

\begin{figure}
\centering
\includegraphics[width=6.5in]{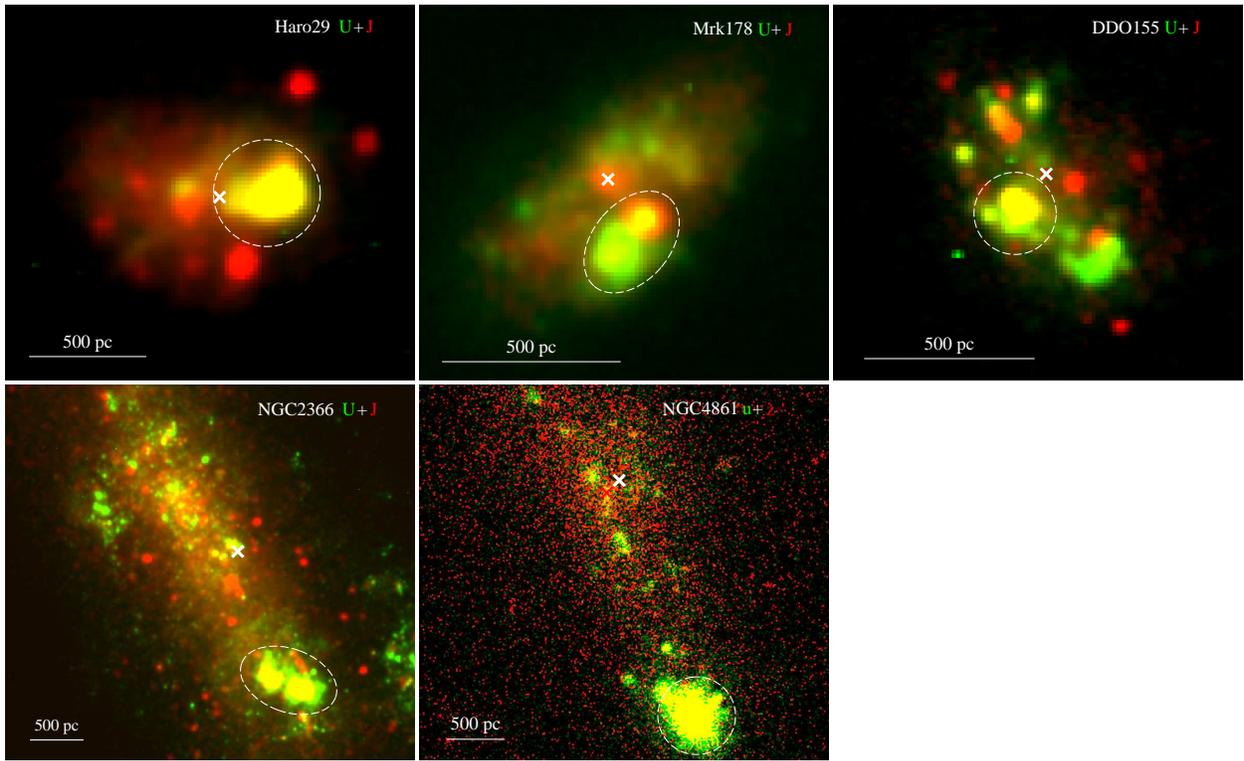}
\caption{Five clumpy irregular galaxies used to study relative clump
mass and possible central migration from tidal torques. The images are
a combination of U and J-band from Hunter \& Elmegreen (2006) except
for NGC 4861, which is a combination of u and z-band from SDSS. The
measured clumps are indicated by elliptical contours and the centers of
the outer V-band isophotes are indicated by ``x''.
}\label{composite}\end{figure}

Many of the observations referenced in the introduction concern BCD or other
dwarf irregular galaxies that have relatively large gas velocity dispersions
compared to the rotation speeds, relatively thick disks compared to the
galactic radii, and relatively massive star-forming regions compared to the
galaxy masses. They are good candidates for the extreme torques and inward
migrations discussed above. Five examples are given in Table \ref{masses},
along with properties of their primary star-formation clumps.  As above, the
clump mass is denoted by $M_{\rm c}$, the galactocentric radius at the center
of the clump is $r_{\rm c}$, and the galaxy mass inside the clump radius is
$M_{\rm dyn}$, measured as $r_{\rm c}v(r_{\rm c})^2/G$ for local rotation
speed $v(r_{\rm c})$. Also for reference, we give the total galaxy stellar
mass, $M_{\rm s}$, the total baryonic galaxy mass (gas+stars), $M_{\rm b}$,
the Coulomb factor $\Lambda\sim(R_{\rm clump}/R_{\rm gal,tot})(R_{\rm
gal,tot}/r_{\rm c})^{1+2\beta}(M_{\rm dyn}/M_{\rm c})$, and the accretion
time $T$ from equation (\ref{time}) for NFW and Burkert dark matter profiles.
The Coulomb factor comes from the approximate expression $\Lambda\sim (R_{\rm
clump}/R_{\rm gal,tot}) (M_{\rm gal,tot}/M_{\rm c})$ in \cite{binney08},
where $R_{\rm clump}$ and $M_{\rm c}$ are the clump radius and mass, and
$R_{\rm gal,tot}$ and $M_{\rm gal,tot}$ are the total galaxy radius and mass.
We take $M_{\rm gal,tot}/M_{\rm dyn}\sim (R_{\rm gal,tot}/r_{\rm
c})^{1+2\beta}$ for rotation curve $v\propto r^\beta$, $R_{\rm clump}$ equal
to half the clump aperture in the table, and $R_{\rm gal,tot}/r_{\rm c}$ from
Figure \ref{composite}. As discussed above, the NFW profile has $\beta=0.5$,
$\xi=0.2$, and a time multiplier of 0.4 to account for faster accretion as
the clump approaches the center; the Burkert profile has $\beta=1$,
$\xi=0.03$, and a multiplier of 0.33.

Color composite images are shown in Figure \ref{composite}.  For four
galaxies, they are made with U and J band images from \cite{hunter06};
for NGC 4861, they are made with u and z band images from the Sloan
Digital Sky Survey (SDSS; Stoughton et al. 2002). The elliptical
contours outline the giant star-forming clumps that we consider to be
candidates for inward migration.  The cross marks the center of the
galaxy as defined by the outer elliptical isophotes in V band.

The HI rotation curves for our galaxies are close to solid body in the clump
region. They are not accurate enough to tell if the dark matter halo is cored
or cuspy (which give rotation curve slopes $\beta=1$ or 0.5 in these two
cases, respectively). To determine $v$ at the clump radius $r_{\rm c}$, we
fit the observed rotation curve to a deprojected speed, $v_{\rm t}$, and
radius, $r_{\rm t}$, at the limit of the observation or the turnover point,
whichever comes first. Then the total galaxy mass inside the clump radius is
taken to be $v_{\rm t}^2r_{\rm c}^3/(r_{\rm t}^2G)$, assuming $\beta=1$ for
this. For the different galaxies, the values of $(v_{\rm t},r_{\rm t})$ in
(km s$^{-1}$, kpc), are, DDO 155: (8, 0.15) \citep{carignan90}; Haro 29: (34,
1.7)\citep{stil02}; NGC 2366: (50, 3.3) \citep{hunter01,thuan04}; NGC 4861:
(40, 3.3) \citep{thuan04}.

For DDO 155 (also known as GR8), the HI observations by \cite{begum03}
suggest that the velocity field is complex, so the dynamical galaxy mass is
inaccurate. \cite{lo93} and \cite{begum03} suggest the velocities have an
expanding or contracting component, which \cite{begum03} fits to a peak value
of 10 km s$^{-1}$, with some radial variations.  They also fit the rotating
part to a peak value of 6 km s$^{-1}$ at the edge (the escape speed was
estimated to be $\sim30$ km s$^{-1}$). The velocity of the giant clump
outlined in Figure \ref{composite} is smaller than the systematic velocity,
so if it is on the near side of the galaxy, then it is expanding away from
the center. \cite{begum03} consider an explosive origin for this motion but
note the lack of old star clusters that might have driven this explosion;
they suggest that HII regions might have had the necessary force. If the
motion is inward, then \cite{begum03} suggest that the clumps might have
coalesced to form the galaxy and are now dispersing to make a disk. They note
that there are no tidal features, however. This inward moving interpretation
is consistent with the model presented in the present paper; tidal features
are not expected because the torques are generated internally. The timescale
for inward motion given in Table 1 is $\sim200$ Myr for DDO 155. This
timescale is consistent with the results of \cite{dohm98}, who find from
resolved stellar population studies that star formation lasts in each clump
for $\sim100$ Myr. They suggest that this long time requires gravitational
self-binding of the clumps.  Dohm-Palmer et al. also suggest that the clumps
come and go on this time scale, with the current generation of clumps at the
positions of the three main HI clouds.  Thus the present model of massive
clump formation by gravitational instabilities in a gas-rich galaxy,
relatively long clump ages from modest gravitational self-binding, and
angular momentum loss through halo, disk, and clump-interaction torques, is
consistent with the HI and stellar observations and previous interpretations
of DDO 155.

Mrk 178 does not have a published rotation curve, but the HI line width
was given by \cite{bott73}, who also derived a total dynamical mass
(``indicative mass'') from the equation $M_{\rm
tot,dyn}=3\times10^{4}r_{\rm H}W^2\;M_\odot$ for Holmberg radius
$r_{\rm H}$ in kpc and linewidth $W$ in km s$^{-1}$ \citep{bott68}.
Scaling to our distance, the Holmberg radius is $r_{\rm H}=2.9$ kpc and
the total dynamical mass is $M_{\rm H}=1.6\times10^9\;M_\odot$.  If we
assume this rotation curve is solid body, then the dynamical mass
inside radius $r_{\rm c}$ is $M_{\rm dyn}(r_{\rm c})=M_{\rm H}(r_{\rm
c}/r_{\rm H})^3$. Setting $r_{\rm c}=0.39$ kpc from Table 1, we get
$M_{\rm dyn}(r_{\rm c})=3.9\times10^6\;M_\odot$.

Clump and galaxy stellar masses were derived by fitting the SEDs over a
range of passbands inside deprojected circular apertures
\citep[see][]{zhang11}. The aperture sizes were determined from the
U-band brightness contours shown in Figure 2. Local background
intensities came from larger annuli around the clumps and were
subtracted from the clump intensities.  For Mrk 178, DDO 155 and NGC
2366, the SEDs used observations in U, B, V, and J passbands
\citep{hunter06}. For Haro 29, we used only U, B and J bands from
\cite{hunter06} because the V-band photometry is inconsistent with the
others. For NGC 4861, we used ugiz data from SDSS.  The giant clump in
NGC 2366 was also studied by \cite{kenn80}. The giant clump in NGC 4861
is known as Mrk 59 and was studied by \cite{izotov09b} and others.

Although there is HI gas present in some of the clumps, and perhaps
even molecular gas connected with current star formation, we do not
include gas in the clump masses. In DDO 155, for example, the HI mass
in the clump is a few times $10^5\;M_\odot$ \citep{carignan90,begum03},
which is comparable to the stellar mass. In NGC 2366, an HI cloud at
the position of the clump contains several$\times10^5\;M_\odot$
\citep{hunter01}, which is $\sim20$\% of the stellar mass. Considering
the possible addition of gas, the clump masses and mass fractions given
in Table \ref{masses} are lower limits.

As part of the fits for clump mass, we also obtained crude star
formation histories in the clumps \citep{zhang11}. These are determined
as relative stellar masses younger than 0.1 Gyr, in the time interval
between 0.1 Gyr and 1 Gyr, and older than 1 Gyr. For the five galaxies,
the relative masses in the intervals ($<0.1$ Gyr, $0.1-1$ Gyr, $>1$
Gyr) are, Mrk 178: (0.42, 0.49, 0.09), DDO 155: ($0.11$, 0.23,
$0.66$), Haro 29: ($0.17$, 0.15, $0.68$), NGC 2366: (0.28, 0.5,
0.22), and NGC 4861: (0.27, 0.50, 0.23). Evidently the SEDs indicate
significant clump components older than 1 Gyr even after background
disk subtraction. The dominant appearance of these clumps in the J-band
(Fig. 2) suggests the same thing. If these old massive components are
really present, then they would have to be gravitationally bound to the
clump and the clump would have to be long-lived.  We note that
gravitational instabilities in a disk of gas and stars can collect both
gas and a significant mass of background field stars into a clump when
the velocity dispersions and densities of the two components are
similar \citep{elmegreen11}. Background field stars also fall into the
clump and get trapped because of the changing gravitational potential
as its mass grows \citep{fellhauer06}.  In our sample, the clumps that
are relatively closest to the center (in Mrk 178, DDO 155 and Haro 29)
contain the highest fraction of old stars. This suggests a larger total
age for the more centralized clumps than for the more peripheral
clumps, which is consistent with a history of inward migration.

The 9th column in Table \ref{masses} gives the ratio between the clump mass
and the galaxy dynamical mass inside the clump radius. As shown in the
previous sections, if this ratio is larger than a few percent, the clump
could significantly perturb the surrounding disk and cold halo particles,
leading to the loss of clump orbital angular momentum in less than $\sim1$
Gyr. The tabulated mass fractions are in this range. The timescale for their
migration is in the last column, assuming NFW and Burkert profiles in two
cases, and using the observed rotation speed at the clump position. The mass
fractions are higher and the timescales are smaller when the clumps are
relatively close to the center, because only the inner parts of the galaxies
are included in $M_{\rm dyn}$.

\subsection{Galaxy Thickness and Scale Length Ratios}

Clump accretion can thicken the central regions because the stellar
orbital energy gets mixed into three dimensions during the final merger
phase \citep{bournaud07}. What is important is the ratio of the disk
scale height, $H=\sigma^2/(\pi G \Sigma)$, to the disk scale length
$R_{\rm d}$. Here, $\sigma$ is the perpendicular velocity dispersion in
the central region of the BCD, and $\Sigma$ is the central mass column
density of the disk. For reference values $\sigma=10$ km s$^{-1}$ and
$\Sigma=10\;M_\odot$ pc$^{-2}$, we obtain $H=740$ pc. Most BCDs in
\cite{hunter06} have $R_{\rm d}\sim500$ pc or less, so $H$ and $R_{\rm
d}$ are comparable. This means the inner parts of BCDs are 3D objects
like a bulge.

Detailed consideration of the BCDs in Table 1 confirm that the inner
disk thicknesses are comparable to or larger than the inner disk scale
lengths. Putting dimensions into the thickness equation, we get
\begin{equation}H=740(\sigma/10\;{\rm km}\;{\rm
s}^{-1})^2(\Sigma/10\;M_\odot\;{\rm pc}^{-2})^{-1} {\rm pc}.
\end{equation}
\cite{zhang11} determined disk stellar mass densities and scale lengths
from SED fits.  Values are given in Table \ref{heights}. The average
$H/R_{\rm d}\sim2.6$, so the BCDs in this study should have relatively
thick inner regions.  The ratio would be larger for larger
perpendicular velocity dispersions -- the assumed value of 10 km
s$^{-1}$ for stars seems to be a lower limit. We note that the BCDs
with giant clumps closest to the center have the highest ratios of
height to length.

Height-to-length ratios greater than unity in Table \ref{heights} are
difficult to understand. They would be smaller if additional mass were
in the disk.  This suggests that some of the BCDs in our survey have a
considerable mass column density of gas in the inner disk, perhaps
comparable to or larger than the stellar column density. Such high
masses of gas might be expected for the clumps in which the starbursts
are occurring (e.g., larger than several hundred $M_\odot$ pc$^{-2}$ in
molecules, which is typical for local giant molecular clouds), but
there might also be a dense molecular and atomic intercloud medium
where the average exceeds the average stellar value of $10\;M_\odot$
pc$^{-2}$. Alternatively, a high filling factor of star-forming gas
clumps that individually have mass column densities in excess of
$\sim100\;M_\odot$ pc$^{-2}$ could produce an average gas column
density in the inner part that exceeds the stellar column density. This
could explain why the BCDs in Table \ref{heights} that have their
massive clumps closest to the center also have the largest
height-to-length ratios, i.e., these galaxies have higher $\Sigma$ than
we assume because of contributions from molecular and dense atomic
material in clumps. Massive clump accretion like that discussed here
would drive significant gas accretion, not only in the clumps but also
of the interstellar material between the clumps, which gets dragged
along with the clumps by gravitational and magnetic forces.

The mass column densities of inner disk HI gas have been observed for
most of these galaxies. For both DDO 155 and NGC 2366, it is
$\sim10\;M_\odot$ pc$^{-2}$ \citep{carignan90,hunter01}. Haro 29 has a
hole in the central HI but the clump is very close and it has an
average column density of $\sim20\;M_\odot$ pc$^{-2}$ \citep{stil02}.
NGC 4861 has a large HI concentration in the center with a column
density of $\sim30\;M_\odot$ pc$^{-2}$ \citep{thuan04}.  Some of these
values are larger than the corresponding central stellar mass column
density by a factor 2 or more, which lowers $H/R_{\rm d}$ in
proportion.  Further studies of the gas column densities in the centers
of BCD galaxies should clarify their relative thicknesses.

\section{Summary}

Young stellar clumps that form by gravitational instabilities in a
galaxy disk can have such a high mass relative to the enclosed galaxy
mass that they produce dynamically significant torques on the halo
stars and cold dark matter particles, on the disk, and on each other.
If the clump mass fraction exceeds a few percent, then these torques
can drive an inflow of the clump's amount of mass in less than 1 Gyr.
This process has been suggested for the formation of bulges in disk
galaxies at high redshift, but it may apply also to local clumpy
galaxies. Because of the general tendency for downsizing, in which
active star formation occurs in galaxies with ever smaller masses as
the universe ages, the clumpy phase now is mostly limited to dwarfs. We
suggest that BCDs are an example of a local clumpy star-bursting galaxy
in which the clumps are large enough to drive significant accretion in
a Gyr or less. This would explain the dense stellar inner disks of
these galaxies, and the prolonged star formation near the center.

The BCDs in our sample also have relatively thick inner regions,
reminiscent of bulges in spiral galaxies. They are even a little too
thick if only the stellar surface densities are considered. This
suggests there could be a dense atomic or molecular component in the
inner region that has an average surface density comparable to or
exceeding that from stars.

This work was funded in part by the National Science Foundation through
grants AST-0707563 and AST-0707426 to DAH and BGE. HZ was partly
supported by NSF of China through grants \#10425313, \#10833006 and
\#10621303 to Professor Yu Gao. We are grateful to the referee for
helpful comments.

\clearpage

\begin{deluxetable}{lcccccccccc}
\tabletypesize{\scriptsize}  \tablecaption{Sample BCD Galaxies and
their Clump Properties\tablenotemark{a}\label{masses}}

\tablehead{ \colhead{Galaxy} &
\colhead{D} &
\colhead{$\log M_{\rm s}$} &
\colhead{$\log M_{\rm b}$} &
\colhead{$\log M_{\rm c}$} &
\colhead{$r_{\rm c}$} &
\colhead{Aperture} &
\colhead{$\log M_{\rm dyn}(r_{\rm c})$} &
\colhead{$M_{\rm c}/M_{\rm dyn}(r_{\rm c})$} &
\colhead{$\ln\Lambda$} &
\colhead{$T$} \\
\colhead{} &
\colhead{Mpc} &
\colhead{$M_\odot$} &
\colhead{$M_\odot$} &
\colhead{$M_\odot$} &
\colhead{kpc} &
\colhead{kpc} &
\colhead{$M_\odot$} &
\colhead{} &
\colhead{} &
\colhead{Gyr}}
\startdata
Mrk 178    & 3.9 & 7.04 & 7.39 & 5.13 & 0.39 & 0.32 & 6.60  & 0.035 & 3.4--4.3 & 0.50--1.4 \\
DDO 155    & 2.2 & 6.47 & 7.22 & 5.46 & 0.21 & 0.24 & 6.78 & 0.048 & 3.2--3.9 & 0.12--0.37 \\
Haro 29    & 5.9 & 7.16 & 8.06 & 6.33 & 0.27 & 0.74 & 6.26 & 1.17  & 1.3--2.3 & 0.03--0.06 \\
NGC 2366   & 3.4 & 7.84 & 9.04 & 6.23 & 1.31 & 0.93 & 8.08 & 0.014 & 3.6--4.0 & 1.3--4.2 \\
NGC 4861   & 7.6 & 8.04 & 8.83 & 6.89 & 2.07 & 0.67 & 8.48 & 0.026 & 1.8--1.8 & 1.7--6.3 \\
\enddata
\tablenotetext{a}{$D$ is the distance, $M_{\rm s}$ is the galaxy stellar
mass, $M_{\rm b}$ is the galaxy baryonic mass, $M_{\rm c}$ is the clump
stellar mass, $r_c$ is the clump galacticentric radius, {\it Aperture} is the
aperture size used for clump photometry, $M_{\rm dyn}$ is the galaxy
dynamical mass inside $r_{\rm c}$, $\Lambda$ is the Coulomb factor, and $T$
is the clump accretion time. For the latter two, we assume $\xi=0.2$ and a
rotation curve slope $\beta=0.5$ in the first case (NFW core), and
$\xi=0.03$, $\beta=1$ in the second case (Burkert core), with factors of
$0.40$ and $0.33$ in $T$, respectively, to account for the decrease in
$M_{\rm dyn}$ with radius.}
\end{deluxetable}

\clearpage

\begin{deluxetable}{lcccc}
\tabletypesize{\scriptsize}  \tablecaption{Inner Scale Heights and
Lengths\tablenotemark{a}\label{heights}}

\tablehead{ \colhead{Galaxy} & \colhead{$\Sigma$} & \colhead{$H$} & \colhead{$R_{\rm d}$} & \colhead{$H/R_{\rm d}$} \\
\colhead{} & \colhead{$M_\odot$ pc$^{-2}$} & \colhead{kpc} &
\colhead{kpc} & \colhead{} } \startdata
Mrk 178    & 4.9 & 1.8 & 0.27 & 6.7 \\
DDO 155    & 7.4 & 1.0 & 0.22 & 4.5 \\
Haro 29    & 39 & 0.19 & 0.20 & 0.95 \\
NGC 2366   & 0.66 & 1.3 & 3.7 & 0.35 \\
NGC 4861   & 10 & 0.74 & 1.0 & 0.74 \\
\enddata
\tablenotetext{a}{$H$ is the inner disk scale height assuming a
perpendicular velocity dispersion of 10 km s$^{-1}$ and the observed
stellar mass column density, $\Sigma$; $R_{\rm d}$ is the inner disk
scalelength. }
\end{deluxetable}

\clearpage

\clearpage

\end{document}